\documentclass[journal]{IEEEtran}

\usepackage{cite}
\usepackage{graphicx}
\usepackage{amsmath}
\usepackage{algorithmic}
\usepackage{array}
\usepackage{stfloats}
\usepackage{url}
\usepackage{color,soul}

\def\BibTeX{{\rm B\kern-.05em{\sc i\kern-.025em b}\kern-.08em
    T\kern-.1667em\lower.7ex\hbox{E}\kern-.125emX}}

\begin{document}

\title{Deep Learning Anomaly Detection for Cellular IoT with Applications in Smart Logistics}

\author{Milos~Savic, Milan~Lukic, Dragan~Danilovic, Zarko~Bodroski, Dragana~Bajovic~\IEEEmembership{Member}, Ivan Mezei~\IEEEmembership{Senior Member}, Dejan Vukobratovic~\IEEEmembership{Senior Member}, Srdjan~Skrbic and Dusan Jakovetic~\IEEEmembership{Member}
\thanks{Milos~Savic, Zarko~Bodroski, Srdjan~Skrbic and Dusan Jakovetic are with Faculty of Sciences, University of Novi Sad, Serbia, e-mail: \{svc, zarko.bodroski, srdjan.skrbic, dusan.jakovetic\}@dmi.uns.ac.rs.

Milan~Lukic, Dragana~Bajovic, Ivan~Mezei and Dejan~Vukobratovic are with the Faculty of Technical Sciences, University of Novi Sad, Novi Sad, Serbia,  e-mail: \{milan\_lukic, dbajovic, imezei, dejanv\}@uns.ac.rs.

Dragan~Danilovic is with the VIP Mobile, Bul. Milutina Milankovica 1z, Belgrade, Serbia, e-mail: d.danilovic@vipmobile.rs.}
\thanks{The work is supported in part by European Commission's Horizon 2020 Research and Innovation Programme, Grant No. 833828.}
}


\maketitle




\begin{abstract}
The number of connected Internet of Things (IoT) devices within cyber-physical infrastructure systems grows at an increasing rate. This poses significant device management and security challenges to current IoT networks. Among several approaches to cope with these challenges, data-based methods rooted in deep learning (DL) are receiving an increased interest. In this paper, motivated by the upcoming surge of 5G IoT connectivity in industrial environments, we propose to integrate a DL-based anomaly detection (AD) as a service into the 3GPP mobile cellular IoT architecture. The proposed architecture embeds autoencoder based anomaly detection modules both at the IoT devices (ADM-EDGE) and in the mobile core network (ADM-FOG), thereby  balancing  between  the  system responsiveness and accuracy. We design, integrate, demonstrate and evaluate a testbed that implements the above service in a real-world deployment integrated within the 3GPP Narrow-Band IoT (NB-IoT) mobile operator network.
\end{abstract}





\section{Introduction}

\IEEEPARstart{T}{he} proliferation of Internet of Things (IoT) and deployment of massive amounts of IoT devices in cyber-physical infrastructure systems such as Smart Factories \cite{DX2014,C+18}, Smart Grids \cite{F+12}, Smart Logistics \cite{T20} and others, brought forward an increasing number of cyber-security \cite{SW+18} and property management challenges \cite{S+18}. For example, Smart Factory or Smart Logistics operations include asset management, intelligent manufacturing, performance optimization and monitoring, planning, and human-machine interaction, but neither of them takes into account full cyber-security protection or data management of Industrial IoT scale~\cite{M+19,H+20}. 
Therefore, handling massive IoT device data integrity and device behaviour in real-time industrial IoT operation and management requires novel approaches. In recent research, they are mainly addressed via various machine-learning (ML) and deep-learning (DL) techniques \cite{Z+19,S+19,Ma+19}. The ability of ML/DL algorithms to process massive data sets while extracting useful features allow them to quickly identify anomalies and prevent breakdowns, which potentially has a broad application space in cyber-physical infrastructures~\cite{CC19,M+20,X+18}.

With the introduction of the 5$^{th}$ generation (5G) cellular networks, IoT cyber-physical infrastructure systems are becoming increasingly reliant on cellular networks \cite{Mu+19}. The 3GPP standardization initiated certain work on support for Cellular IoT (CIoT) during the 4G Long-Term Evolution (4G LTE) development \cite{R1}, which resulted in first CIoT technologies such as Narrow-Band IoT (NB-IoT) that has been introduced in 
the 3GPP Release 13 \cite{R2,R+20}. This work has since then expanded to Ultra-Reliable Low-Latency Communications (URLLC) and massive Machine-Type Communications (mMTC) services in 5G \cite{P+16}. As billions of new CIoT devices are expected to be connected world-wide in the following years, providing efficient and automated monitoring and threat detection both at the CIoT devices and within the CIoT network architecture will be critical to securely manage devices and cover this attack surface \cite{B+17,Z+17}.

In this paper, we propose to augment the 3GPP mobile cellular architecture with additional enhancements that provide support for a network-wide anomaly detection (AD) service. Our target is a generic AD CIoT service which can be tailored to applications ranging from identifying malfunctioning devices to threat detection for secure CIoT. The proposed hierarchical AD architecture embeds anomaly detection modules (ADMs) both at the IoT devices (ADM-EDGE) and in the mobile core network (ADM-FOG). The ADM modules are based on both shallow and deep autoencoders (AE) whose complexity is matched to both the edge and the fog deployment, balancing between the system responsiveness and accuracy. The distinguishing feature of our work is that the proposed AD enhancement of the CIoT architecture, including both ADM-EDGE and ADM-FOG modules, is implemented and deployed in a real-world CIoT network based on the 3GPP NB-IoT standard and demonstrated in the context of Smart Logistics. Moreover, we custom-designed a novel NB-IoT device platform for a Smart Logistics use case, where NB-IoT devices are connected to shipping containers in a factory supply chain, in order to collect data, deploy and test the ADM-EDGE module.

The paper is organized as follows. In Sec. \ref{Sec2}, we provide technical background, review the related work and present the contributions of this paper. The proposed solution for DL-based anomaly detection in CIoT is presented in detail in Sec. \ref{Sec3}. In Sec. \ref{Sec4}, we describe system integration, data generation and provide numerical results from real-world experiments. The paper is concluded in Sec. \ref{Sec5}.

\section{Background}
\label{Sec2}

In this work, we augment the CIoT architecture with anomaly detection capabilities at the IoT devices (edge) and the mobile core network servers (fog). Before going to details, we first provide the technical background needed for understanding the proposed system architecture and functionalities.

\subsection{3GPP Cellular IoT Architecture}

\begin{figure*}[t]
\centerline{\includegraphics[width=5.6in,height=2.5in]{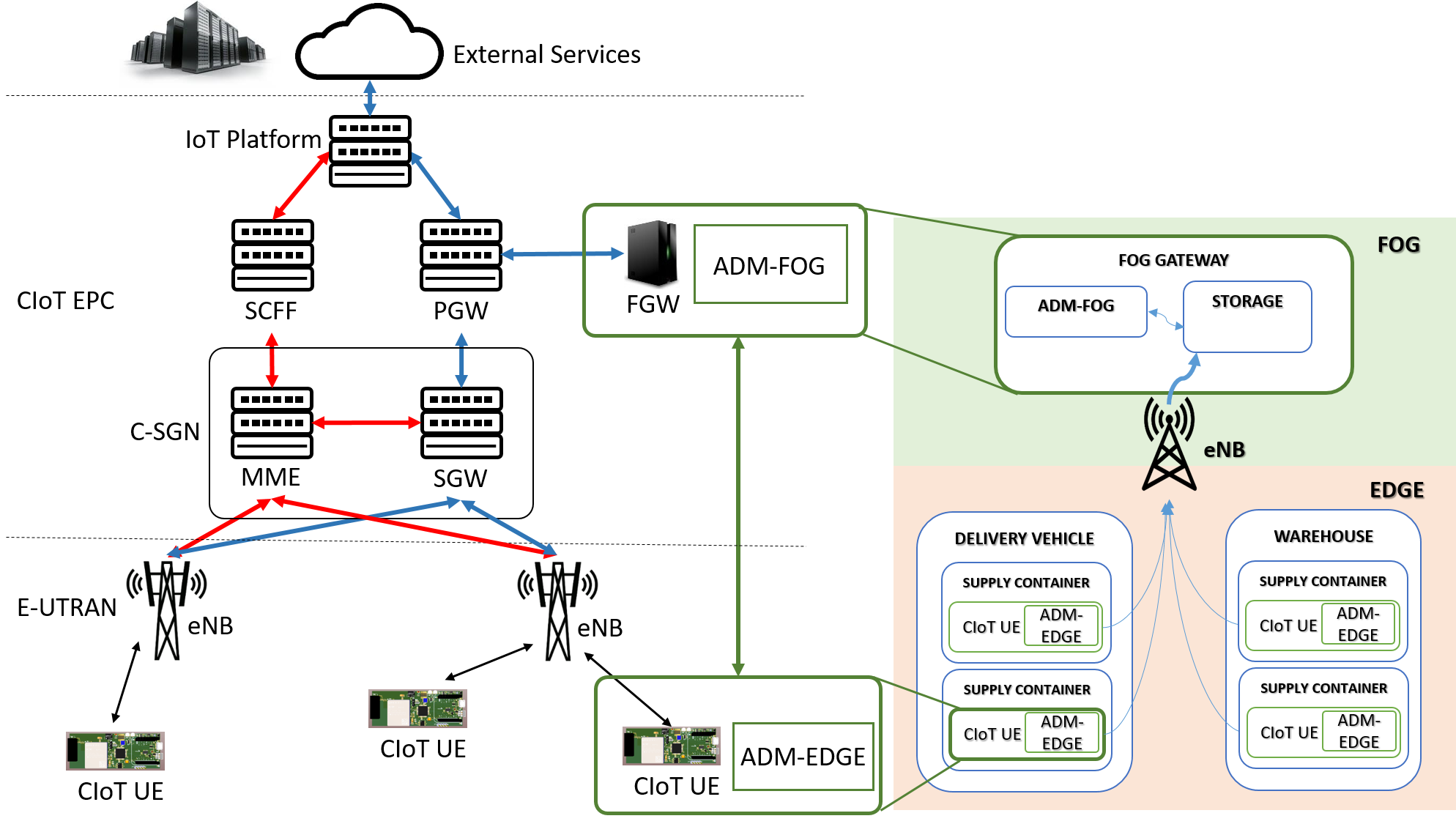}}
\caption{3GPP CIoT architecture augmented with Anomaly Detection enhancements.}
\label{Fig_1}
\end{figure*}

We start by describing the current state-of-the-art CIoT architecture focusing primarily on the 3GPP NB-IoT technology~\cite{R1,R2}. NB-IoT is a new CIoT technology that can be seamlessly integrated in the existing 3GPP 4G/5G architecture, coexisting in the radio access network with the current 3GPP 4G LTE and the emerging 3GPP 5G NR technology, and using the same evolved packet core (EPC) network functionalities \cite{W+17}. Focusing on the current 3GPP 4G LTE architecture, the relevant 3GPP CIoT architecture elements are illustrated in Fig. \ref{Fig_1}. CIoT user equipment (CIoT UE), which is a formal name for a NB-IoT device, connects to the network via a neighbouring base station or eNodeB (eNB), which is the main element of the Evolved Universal Terrestrial Radio Access Network (E-UTRAN). NB-IoT downlink/uplink resources are allocated either within 4G LTE band (in-band deployment), at its edge (guard-band deployment), or as a separate channel (out-of-band deployment). After eNB, both user-plane (i.e., user data packets) and control-plane (i.e., signalling messages) information is processed at a CIoT Serving Gateway Node (C-SGN), which covers functionalities of both control-plane Mobility Management Entity (MME) and user-plane Serving Gateway (SGW). User-plane data further flows through a Packet Gateway (PGW) to the IoT platform, which forwards data via the Internet to external network application servers \cite{book-nbiot}.

Two options for data transfer between the CIoT UE and the IoT platform are envisioned. The first one (mandatory) uses signalling radio bearers to transmit user data, thus avoiding establishment of data radio bearers for energy efficiency. From eNB, data is routed either following a control-plane path via an EPC element called Service Capability Exposure Function (SCEF) for non-IP data, or a user-plane path via C-SGN and PGW for both IP/non-IP data. The second one (optional) establishes a data radio bearer to send IP/non-IP data via an eNB/C-SGN/PGW user-plane path to the IoT platform. Herein, we assume that a UDP encapsulated IP data from CIoT UE device traverses the path following the latter approach, which will impact the deployment choices for the proposed anomaly detection enhancements strategy described in Sec. \ref{Sec3}.

\subsection{Machine Learning for Anomaly Detection at the Edge}

Security challenges and threats in industrial IoT networks call for innovative applications of ML/DL techniques for IoT security. More specifically, these techniques can be employed for authentication and access control, anomaly and intrusion detection, malware analysis and distributed denial-of-service (DDoS) attacks detection and mitigation \cite{hhhh,U+19}. The main challenges of implementing ML/DL models at the edge are scalability issues and IoT edge platforms resource limitations \cite{M+20}. Depending on the ML algorithm being run on the edge node, the size of the ML model can go as low as a few kilobytes. Also, the requirements in regard to the memory capacity and computational power depend heavily on the choice whether the models are trained at the edge, or pre-trained models are being used.

Besides the sensor readouts, which are the primary source of data for ML/DL at the edge, an IoT module itself can provide a host of useful insights about the network and wireless link conditions, the feature we also exploit in our edge device design described in Sec. \ref{edge_device_design}. The amount of useful data that can be extracted from the IoT module generally exceeds the capacity of the wireless communication channel, however, this kind of metadata can be used to feed a locally run ML algorithm for anomaly detection, or be aggregated and sent to the core network fog gateway periodically, for further analysis.

In this work, to perform AD, we apply shallow and deep autoencoders (AE) trained using deep learning algorithms. AE is a neural network that learns a latent lower-dimensional representation of training data by reproducing its inputs through latent variables in the hidden layers at the output layer with the smallest possible error. The error function captures differences between values at the input and output layers. This so-called reconstruction error is used as the outlier score in an anomaly detection process. The proposed AD architecture is hierarchical, as it comprises AD models running at different levels within a CIoT system (both IoT edge devices and core network fog gateway), where more powerful higher-level models are activated if decisions of lower-level models have low confidence scores (see Sec.~\ref{ADSubSection} for details).

\subsection{Related Work}

Recent research efforts in the area of ML methods for anomaly detection at the edge IoT devices have been focused on efficient utilization of limited computational resources at the edge. It is well-known that the training process for most of deep learning-based AI models is highly resource-intensive, usually requiring hardware resources (e.g., GPU, FPGA)~\cite{zzxc}. Resource-aware edge AI model designs have been considered in a different line of research. The AutoML idea~\cite{yhjl} and the Neural Architecture Search techniques \cite{zqvl} have been used to devise resource-efficient edge AI models tailored to the hardware resource constraints of both the underlying edge devices and network servers. Important research advances were also made regarding the tailored design of DL architectures for resource-constrained devices: Zhang et al. proposed an extremely efficient convolutional neural network (CNN) for mobile devices and Nikouei et al. introduced a lightweight CNN that can run on edge devices~\cite{mmln}.

A number of proposals using distributed ML/DL for security in Industrial IoT are recently considered \cite{T+19}. In DIoT, a recurrent neural network (RNN) is trained for each device type present in the IoT network to learn a normal communication profile. A federated (distributed) learning scheme is employed to learn device-type specific RNNs~\cite{ndfs}. Wang et al. proposed a control algorithm that determines the best trade-off between local update and global parameter aggregation in data partitioned federated learning models trained using gradient-descent algorithms~\cite{wafl}. Ferdowsi and Saad proposed a distributed privacy preserving IoT intrusion detection security system based on federated generative adversarial networks. In the proposed decentralized architecture, every IoT device monitors its own data as well as neighbor IoT devices to detect internal and external attacks~\cite{fsga}. Meidan et al. proposed N-BaIoT -- a method for detecting IoT botnet attacks based on deep autoencoders. For each device present in an IoT network, a deep autoencoder is trained on features extracted from normal traffic data~\cite{mnbn}. Bezerra et al. proposed IoTDS -- a distributed method for detecting IoT botnet attacks based on light-weight one-class classification models~\cite{bito}. Rathore and Park created a decentralized attack detection framework for IoT networks based on semi-supervised learning employing extreme learning machines and fuzzy C-means algorithms~\cite{rpss}. Doshi et al. employed various machine learning algorithms (k-nearest neighbor, support vector machines, decision trees and neural networks) to detect DDoS attack traffic in consumer IoT devices~\cite{dmld}. Pajouh et al. (2018) proposed a malware detection approach for IoT based on deep RNNs~\cite{pdrn}, while
\cite{jsal} presents an approach to anomaly detection that implements autoencoders at each edge device,  while the edge devices are orchestrated via a federated learning model with the central server.
In \cite{kkbd}, the authors show that Random Forest, Multilayer Perceptron, and Discriminant Analysis models can viably save time and energy on the edge device during data transmission, while K-Nearest Neighbors, although reliable in terms of prediction accuracy, is resource-inefficient in their studies.

\begin{figure*}[t]
\centerline{\includegraphics[width=5.7in,height=1.8in]{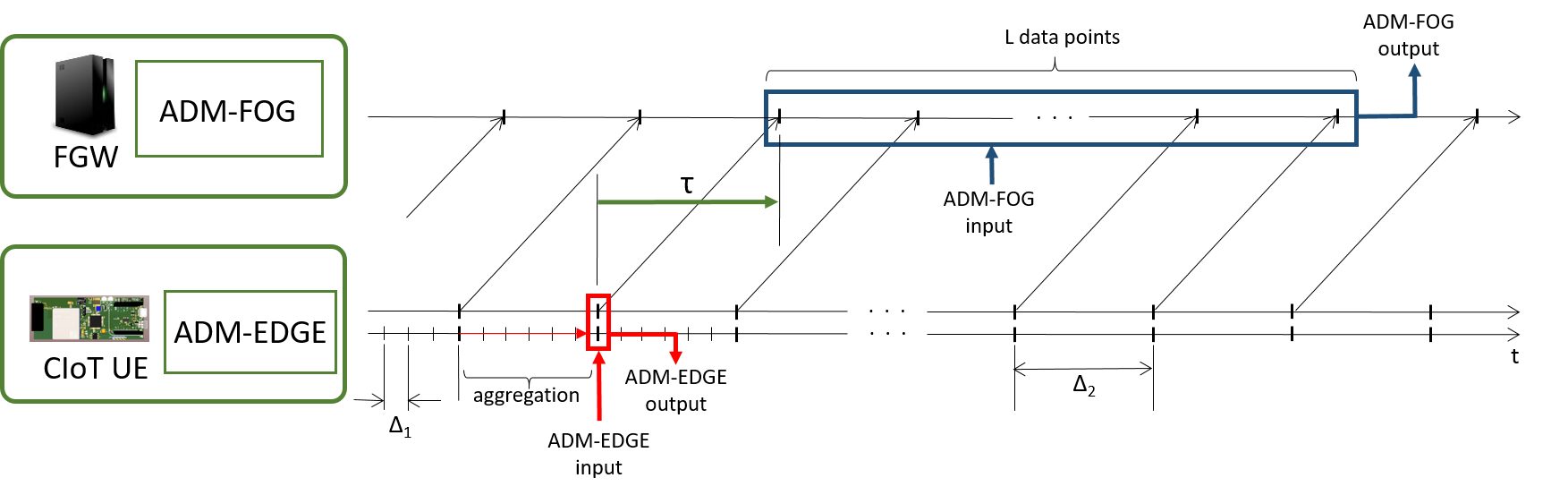}}
\caption{3GPP CIoT Anomaly Detection processing flow.}
\label{Fig_2}
\end{figure*}

\subsection{Contributions}

We now summarize the main contributions of the paper. We propose an approach to embed anomaly detection capabilities in the Cellular IoT architecture, providing for combined threat detection both at the IoT devices (edge) and in the mobile core network servers (fog). The corresponding architecture design is motivated by and well-suited for Smart Logistics. The proposed edge-based ADM-EDGE and fog-based ADM-FOG modules can balance between the responsiveness and accuracy by employing both shallow and deep autoencoder (AE) based learning modules whose complexity is matched to both edge and fog deployment. We carry out implementation, integration, and evaluation of an end-to-end testbed according to the proposed architecture. This includes: 1) real IoT data generation and emulation of a real-world Smart Logistics scenario; 2) fabrication and configuration of the relevant edge and fog hardware and infrastructure; 3) development and implementation of a software library for edge and fog-based anomaly detection; and 4) evaluation of the developed anomaly detectors on the generated data and quantification of detection performance-response time\footnote{Response time is the time passed from the occurrence of an anomaly to its detection.} tradeoffs. For the latter contribution, we explicitly quantify the tradeoffs that take into account limited computational and storage budget at the edge devices, and communication and processing costs due to processing larger amounts of data at the fog for improved AD performance.

\section{DL-Based Anomaly Detection in 3GPP NB-IoT}
\label{Sec3}

In this section, we describe in detail the design and system architecture of the proposed AD support for the 3GPP NB-IoT mobile cellular network.

\subsection{System Model and Architecture}

We augment the 3GPP CIoT system architecture with support for CIoT device anomaly detection. The augmented architecture is illustrated in Fig. \ref{Fig_1} and introduces two additional ADMs: one placed at the edge CIoT UE (ADM-EDGE) and another placed at the fog gateway (ADM-FOG). The architecture represents a generic CIoT enhancement for anomaly detection, although in this work, we specialize it to the domain of Smart Logistics. This includes managing supply of items from various origin points delivered to warehouses in manufacturing plants (Fig.~\ref{Fig_1}). Items being delivered are packed into containers, each of which has an NB-IoT device attached. For this purpose, we designed an entirely new NB-IoT UE device, and deployed suitable ADM-EDGE and ADM-FOG modules at both NB-IoT UEs and the FGW server within the mobile core network.

\textbf{ADM-EDGE:} As described below, NB-IoT devices collect various information such as acceleration and GPS coordinates. This sensory information can be used to detect anomalies such as physical tampering of items, container mishandling such as overturning, delays, routing problems, incidents with the delivery vehicles, etc. We assume each NB-IoT device possesses two types of sensors: i) sensor S1  with low sampling rate $f_1$ [Hz] and sampling period $\Delta_1=\frac{1}{f_1}$ [s] (we consider a GPS sensor that samples the outdoor device location), and 2) sensor S2 with high sampling rate $f_2$ [Hz] and sampling period $\Delta_2=\frac{1}{f_2}$ [s] (we consider accelerometer/gyroscope that samples vibration monitoring parameters), as illustrated in Fig. \ref{Fig_2}.

Due to a limited memory capacity and processing power, ADM-EDGE integrated into an NB-IoT device firmware requires restrictive design. ADM-EDGE consists of a pre-trained autoencoder detecting anomalies in individual data points. At the input, ADM-EDGE processes a single data point that consists of a single S1 and S2 value. As illustrated in Fig. \ref{Fig_2}, we assume ADM-EDGE is triggered synchronously with the low-rate sensor S1 outputs $X_{S1}[k] = X_{S1}(t = k\Delta_1), k=\{1,2,\ldots\}$, where $\Delta_1$ is the sampling period of the sensor S1 output function $X_{S1}(t)$. Besides an S1 sample, ADM-EDGE is fed with the sensor S2 value $X_{S2}[k]$, which is a root mean square (RMS) aggregate value of high-rate sensor S2 output samples calculated over the interval of duration $\Delta_1$ between the last two S1 outputs. In other words, $X_{S2}[k]=\sqrt{\frac{1}{M}\sum_{\ell} X_{S2}^2(t = \ell\Delta_2)}$, where $\ell$ satisfies $(k-1)\Delta_1 < \ell\Delta_2 \leq k\Delta_1$, which amounts to the last $M = \frac{\Delta_1}{\Delta_2}$ S2 samples preceding $t=k\Delta_1$. To summarize, a pair of S1 and aggregated S2 values $(X_{S1}[k],X_{S2}[k])$ represents a data point fed into an ADM-EDGE autoencoder every $\Delta_1$[s]. For each decision, after he current ADM-EDGE processing is completed, the device outputs a confidence score (see Sec.~\ref{ADSubSection}).

\textbf{ADM-FOG:} NB-IoT devices connect to a mobile network and transfer data via the nearest base station. Each ADM-EDGE data point is forwarded to the FGW, adjoined with the ADM-EDGE confidence score evaluated from the last available data point. The communication delay incurred by NB-IoT network connection may vary between an order of tens-of-milliseconds to several tens-of-seconds, depending on the NB-IoT device radio conditions and network load \cite{Mar+19}. The FGW server runs an instance of ADM-FOG relying on higher memory capacity and processing power. Thus, ADM-FOG uses a more powerful autoencoder that processes multi-variate timeseries through several hidden layers. At ADM-FOG, a larger input is considered which is formed by concatenating the last $L$ ADM-EDGE data points (see Fig. \ref{Fig_2}). Thus at the time instant $t_k$ when the $k$-th data point is received at the FGW (note that $t_k = k\Delta_1 + \tau_k$, where $\tau_k$ is communication delay of the $k$-th data point), ADM-FOG is triggered with the input containing the set of the last $L$ data points $\{(X_{S1}[i],X_{S2}[i])\}_{k-L<i\leq k}$ received prior to the time instant $t_k$. Decisions made by ADM-EDGE are revised in case that the corresponding confidence scores take values below a certain threshold.

To summarize, the above AD-augmented CIoT architecture features several important properties: 1) ADM-EDGE at the NB-IoT node immediately detects an anomaly over a single data point which may result in extremely fast response time (order of milliseconds); 2) ADM-FOG collects timeseries of specific lengths matched to the more powerful AE design through a communication channel that can be a bottleneck and cause unpredictable delays (order of seconds); 3) Only ADM-EDGE has access to raw data (note that sending the raw data to ADM-FOG would be inefficient due to a low-rate NB-IoT connection and energy-constrained NB-IoT devices), while ADM-FOG gets access to aggregated data; 4) ADM-FOG applies deep learning analyses over the longer timeseries of data points using more powerful AE design with more hidden layers, requiring higher processing power and memory capacity unavailable at the edge; 5) In the worst case scenario, the final anomaly detection decision at the system level is obtained within the time frame of several seconds. It is worth noting that this response time meets the requirements and is well-aligned with the targeted Smart Logistics applications \cite{r42}.

\subsection{NB-IoT Edge Device Design}
\label{edge_device_design}

\begin{figure}[ht]
\centerline{\includegraphics[width=3.3in]{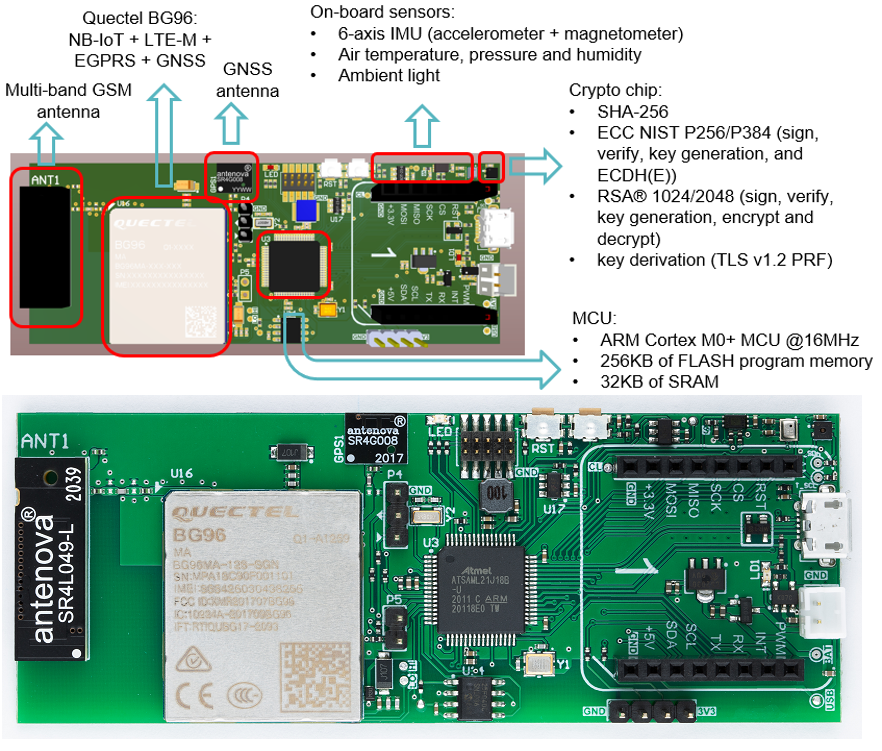}}
\caption{3GPP NB-IoT/LTE-M edge node running ADM-EDGE model.}
\label{Fig_Edge_node}
\end{figure}

We designed the NB-IoT edge device illustrated in Fig. \ref{Fig_Edge_node} having in mind the specific requirements of a Smart Logistics environment: tracking and monitoring the vibration of the shipping containers. Here, we reflect on the most important features supported by our device.

\subsubsection{Cellular connectivity}

To fulfill the requirement for ubiquitous connectivity, while keeping the power consumption of the battery-powered device low, we utilize a BG96 cellular module from Quectel, which supports NB-IoT and LTE-M, as state-of-the-art 3GPP CIoT communication standards, that will be further evolved in 5G standardization \cite{Ratasuk_2019}. In addition, EGPRS is supported to ensure the connectivity in areas where LTE carrier might not be available. Finally, the integrated GNSS module provides the geolocation information which is essential to the asset tracking task in the logistics use case. The intention is to use NB-IoT as the primary means of communication due to its desirable properties, namely energy efficiency combined with extended coverage \cite{V+17}. However, in occasions when it is necessary to transfer larger amounts of data, (e.g. a new firmware image), LTE-M is a more efficient solution. The architecture of our edge node provides flexibility which allows us to adapt the throughput of the communication module according to the needs of the application.

\subsubsection{On-board sensors}
Apart from the localization data provided by the GNSS module, on-board environmental sensors are used to measure parameters relevant to the logistics use case. The 6-axis Inertial Measurement Unit (IMU) provides information about the vibrations and the magnetic field along X, Y and Z axes relative to the chip position. An additional set of sensors is used to measure the atmospheric conditions such as air temperature, pressure and humidity.

The designed platform provides additional metadata that could be used as inputs to ADM-EDGE. For example, the cellular modem is capable of providing the standard set of radio condition metrics (SNR, RSSI, RSRP, etc.). In addition, our design includes the on-board current measuring circuitry that allows the micro-controller unit (MCU) to acquire precise measurements of power consumption by the BG96 module.

\subsubsection{The MCU features and capabilities}

The main MCU inside the edge node is a low-power 32-bit ARM Cortex M0+ with 256KB of FLASH and 32KB of SRAM, operating at 16MHz. The MCU resources are sufficient to efficiently control the rest of the circuitry, while maintaining low power consumption, especially in the sleep mode. However, the absence of operating system as well as the hardware constraints limit the usage of ML tools only to lightweight models that are fully customized and optimized for a given application. Finally, an external FLASH memory module enables data logging over the intervals when there is no connectivity, and it is used to store the firmware images during over-the-air updates.

\subsubsection{Security}

In an industrial setup, the security is of the critical importance. Thereby, we use a hardware crypto element which enables offloading the computationally expensive asymmetric cryptographic algorithms (elliptic-curve cryptography and RSA) from the resource-constrained MCU \cite{HWSpaper}. Tampering-resistant memory within the crypto chip is used to store security credentials, making FW on the host MCU oblivious of the sensitive information such as the encryption keys and certificates.

\subsection{Anomaly Detection using ADM-EDGE and ADM-FOG}
\label{ADSubSection}

\subsubsection{Autoencoder inference and training}

ADM-EDGE and ADM-FOG detect anomalies using autoencoders. An autoencoder is a feed-forward neural network trained to replicate input values at the output layer in order to obtain latent data representations in hidden layers. The number of neurons in the output layer of the autoencoder is equal to the number of neurons in the input layer, and both quantities are equal to the number of features in the training dataset (the $i$-th neuron in the input/output layer represents the $i$-th feature from the training dataset). The number of neurons in hidden layers has to be significantly smaller than the number of features in order to avoid learning autoencoders realizing trivial linear identity functions and to obtain useful lower-dimensional data representations capturing the most salient latent features.

\begin{figure*}[htb!]
\centerline{\includegraphics[width=5in]{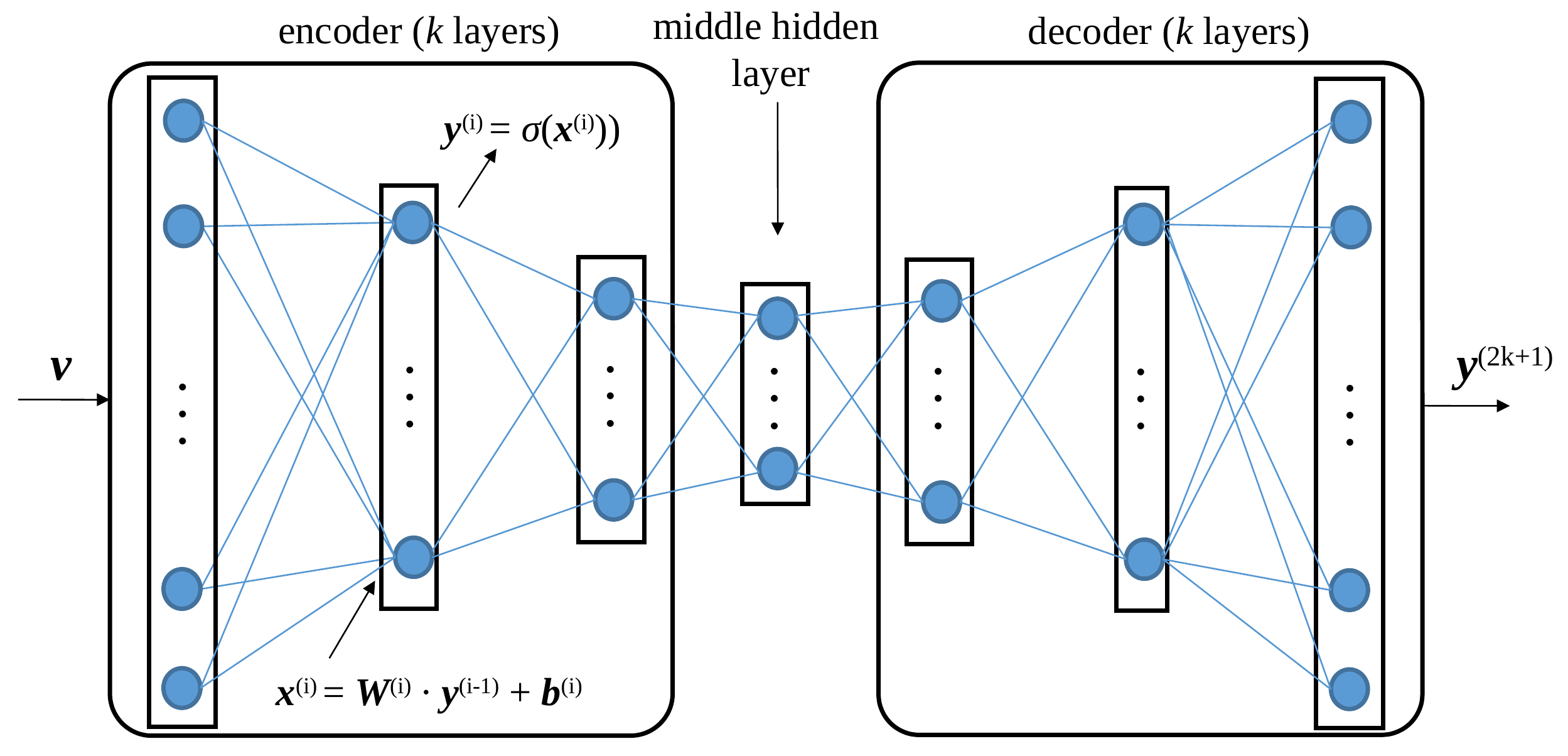}}
\caption{Autoencoder architecture.}
\label{Fig_AE_structure}
\end{figure*}

Autoencoders typically have a symmetric architecture with an odd number of layers as shown in Figure~\ref{Fig_AE_structure}. The first $k$ layers, each having a smaller number of nodes than the previous layer, represent an encoder function producing a lower-dimensional data representation in the $(k+1)$-th layer (the middle hidden layer). The next $k$ layers constitute a decoder function reconstructing original feature values from latent features learned in the middle hidden layer. In contrast to encoder layers, the number of nodes in decoder layers increases with each next layer so that the $d$-th decoder layer has the same number of nodes as the $(k + 1 - d)$-th encoder layer for $d \in [1 \: .. \: k]$. 
For ADM-EDGE and ADM-FOG autoencoders we consider the following architectures:
\begin{enumerate}
\item autoencoder architecture with 1 hidden layer in which the middle hidden layer contains $n/2$ nodes, where $n$ is the total number of features,
\item autoencoder architecture with 3 hidden layers containing sequentially $n/2$, $n/4$ and $n/2$ nodes, and
\item autoencoder architecture with 5 hidden layers containing sequentially $3n/4$, $n/2$, $n/4$, $n/2$ and $3n/4$ nodes.
\end{enumerate}
As it will be explained later, the principal difference between ADM-EDGE and ADM-FOG autoencoders is not related to their architecture, but to the type of input they are accepting and processing.


As for any feed-forward neural network, there is a directed weighted link from each node in the $i$-th layer to each node in the $(i + 1)$-th layer. During the autoencoder inference process (i.e., when applying the autoencoder to a vector of input values), a node in the $(i + 1)$-th layer accumulates outputs from the nodes in the $i$-th layer that were previously multiplied by link weights, optionally adding a bias value to the accumulated result. Then, an activation function is applied to the accumulated result to form the output value of the node. Let $v$ denote a vector of input values for the autoencoder where $v_{i}$ is the value of $i$-th feature in $v$. With $x^{(l)}$ and $y^{(l)}$ we denote the vector of input and output values for layer $l$, respectively, where $x_{i}^{(l)}$ and $y_{i}^{(l)}$ are the input and the output value of the $i$-the node in layer $l$, respectively.
The feed-forward autoencoder inference process can be formally described by the following equations:
\begin{equation}
\label{autoencoder_feed_forward}
\begin{array}{r@{}l}
    x_{i}^{(l)} &{} = W_{i}^{(l)} \cdot y^{(l - 1)} + b_{i}^{(l)}\\
    y_{i}^{(l)} &{} = \sigma(x_{i}^{(l)})\\
    y_{i}^{(1)} &{} = v_{i},
\end{array}
\end{equation}
for each node $i$ in each layer $l$ ($l > 1$). Here, $W_{i}^{(l)}$ is the vector containing weights of autoencoder links pointing to the $i$-th node in layer $l$ and $b_{i}^{(l)}$ is the bias value associated to the same node; $W_{i}^{(l)} \cdot y^{(l - 1)}$ is the scalar product of the previously mentioned vector of weights and the vector of outputs from nodes contained in the previous layer; and $\sigma(x)$ stands for the activation function. ADM-EDGE and ADM-FOG autoencoders use the ReLU (Rectified Linear Unit) activation function: $\sigma(x) = \text{max}(0, x)$.

The weights of autoencoder links and biases associated to autoencoder nodes are learnt on a training dataset $T$ by minimizing an error (or loss) function $E$. Let $\theta$ denote the set of autoencoder trainable parameters (weights and biases) and let $\mathcal{A}(v, \theta)$ be the output of the autoencoder for the vector of input values $v$ (values in the last layer after the feed-forward inference procedure). ADM-EDGE and ADM-FOG autoencoders are trained by minimizing the mean squared error (MSE) function on $T$:
\begin{equation}
\label{MSE}
E(T, \theta) = \frac{1}{|T|}\sum_{v \in T} \text{Err}(v, \theta),
\end{equation}
where $|T|$ is the number of instances in $T$. $\text{Err}(v, \theta)$ is the reconstruction error of instance $v$:
\begin{equation}
\label{Err}
\begin{array}{r@{}l}
\mbox{Err}(v, \theta) &{}= \sum_{i = 1}^{f} (v_{i} - \hat{v}_{i})^{2}\\
\hat{v}       &{}= \mathcal{A}(v, \theta),
\end{array}
\end{equation}
where $f$ is the number of features in $T$.

To minimize MSE of ADM-EDGE and ADM-FOG autoencoders we use the Adam optimization algorithm~\cite{adam2014}. Adam belongs to the class of iterative gradient descent (GD) optimization algorithms which minimize $E(T, \theta)$ by updating $\theta$ in the opposite direction of the gradient of $E(T, \theta)$ with respect to $\theta$, denoted by $\nabla_{\theta}E(T, \theta_{t})$. There are three variants of GD considering the number of data instances that are used to compute the gradient and update $\theta$: (1) the batch GD uses the whole training dataset $T$, (2) the mini-batch GD performs updates on subsets on $T$, and (3) the stochastic GD updates $\theta$ on individual data instances. The mini-batch approach is employed to train ADM autoencoders. The Adam algorithm iteratively updates $\theta$ by the following rule:
\begin{equation}
\label{ADAM_update}
\theta_{t} = \theta_{t - 1} - \eta \frac{\hat{m_{t}}}{\sqrt{\hat{v_{t}}} + \epsilon},
\end{equation}
where $\eta$ is the learning rate (by default $\eta = 0.001$), $\hat{m_{t}}$ and $\hat{v_{t}}$ are bias-corrected first and second moment (mean and variance) estimates of $\nabla_{\theta}E(T^{b}, \theta_{t})$, where $T^{b}$ is a mini-batch of $T$, and $\epsilon$ is a small value to prevent division by zero ($\epsilon = 10^{-8}$). $\theta_{0}$ is initialized by the uniform initializer proposed by Glorot and Bengio~\cite{Glorot2010}.
Bias-corrected moment estimates are derived from raw moment estimates $m_{t}$ and $v_{t}$ initially set to zero vectors:
\begin{equation}
\label{Adam_moment_estimates}
\begin{array}{r@{}l}
\hat{m_{t}} &{}= m_{t} \: / \: (1 - \beta_{1}^{t}) \\
\hat{v_{t}} &{}= v_{t} \: / \: (1 - \beta_{2}^{t}) \\
m_{t}       &{}= \beta_{1} m_{t - 1} + (1 - \beta_{1}) g_{t} \\
v_{t}       &{}= \beta_{2} v_{t - 1} + (1 - \beta_{2}) g_{t}^{2} \\
g_{t}       &{}= \nabla_{\theta}E(T^{b}, \theta_{t - 1}).
\end{array}
\end{equation}
Hyper-parameters $\beta_{1}$ and $\beta_{2}$ control decay rates of mean and variance estimates of the gradient, respectively. Default values are $\beta_{1} = 0.9$ and $\beta_{2} = 0.999$. $\beta_{i}^{t}$ is $\beta_{i}$ to the power $t$ (for $i = 1, 2$).

\subsubsection{Anomaly detection based on autoencoders}

Let us assume that the device behaviour is described by a feature vector $X$ containing $k$ real-valued features. Those may be values of sensory readings observed at one particular point in time or multivariate timeseries of consecutive sensory readings. Let $D$ denote a set of data points that depicts the normal (nominal) behaviour of the device (the training dataset), let $\mathcal{A}$ be an autoencoder trained on $D$, and let $e$ denote the maximal reconstruction error on $D$:
\begin{equation}
\label{instance_error}
e = \max_{v \in D} \mbox{Err}(v, \theta).
\end{equation}
Then, a data point $y$ not contained in $D$ (a data point that is not present in the training dataset) can be considered as an anomaly if the reconstruction error of $y$ by $\mathcal{A}$ is higher than $e$, i.e. $\mbox{Err}(y, \theta) > e$. In other words, a data point is not an anomaly if it is better reconstructed by $\mathcal{A}$ than the worst reconstructed data point from the training dataset.

ADM-EDGE and ADM-FOG autoencoders identify anomalies according to the previously described rule. For each anomaly detection decision, the confidence score $C(y)$ is computed according to the following formula:
\begin{equation}
\label{formula_cs}
C(y) = S(\mbox{Err}(y,\theta) - e),
\end{equation}
where $S$ denotes the sigmoid function, $S(x) = 1 / (1 + e^{-x})$.

The important property of the confidence score function is that non-anomalous data points have scores in the range (0,~0.5], whereas anomalous data points exhibit higher scores that belong to the interval (0.5, 1). In other words, confidence scores close to 0 indicate non-anomalous data points, while values close to 1 signify anomalies. Thus, confidence scores for non-anomalous data points after making decision are further transformed into $1 - C$, where $C$ is a value obtained by Eq.~\eqref{formula_cs}.

Due to a low computational power and small memory capacity, it is practically infeasible to train the ADM-EDGE autoencoder directly on the edge node device: 1) a large number of data points has to be stored at the device to train a model exhibiting an acceptable level of accuracy, 2) the training of autoencoders is a computationally intensive optimization process usually performed in a large number of iterative steps, 3) a low computational power prevents any serious model validation and tuning of model hyper-parameters. Consequently, we adopt a scheme in which ADM-EDGE autoencoders are trained offline and an inference engine for feed-forward neural networks is directly integrated into the firmware of the edge node device enabling autoencoder-based anomaly detection on pretrained models. The edge node device also does not provide a storage for sensory readings. This means that it is also not feasible to make ADM-EDGE autoencoders detecting anomalies in timeseries data. Thus, ADM-EDGE autoencoders perform anomaly detection considering individual data points (the last values of sensory readings). In our future work we will also consider online training for ADM-EDGE autoencoders for more powerful edge node devices w.r.t. computational and storage capabilities.
In contrast to ADM-EDGE lightweight autoencoders, ADM-FOG autoencoders process multivariate timeseries constructed using the sliding window approach in an arbitrary number of hidden layers.

The inference engine for ADM-EDGE autoencoders is realized in C as a standalone, self-contained module without any external dependencies to third party libraries.
This C module is directly integrated into the firmware of an edge node device.
To train ADM-EDGE and ADM-FOG autoencoders we have developed a Python module based on the deep learning Tensorflow library~\cite{TENSORFLOW}. This module builds an autoencoder as a Tensorflow sequential neural network model for a given specification of the autoencoder structure (the number of hidden layers and the number of nodes per hidden layer) and determines autonecoder weights and biases using the previously described Adam algorithm for a given number of epochs (by default 100) and batch size (by default 16). Before training, data points in the input training dataset are normalized such that each feature has zero mean and unit variance. The traininig of both ADM-EDGE and ADM-FOG autoencoders is performed on a fog gateway. In the case of ADM-EDGE autoencoders, the structure of the trained autoencoder AD model, its weights and data normalization parameters are exported as C declarations to a header file. The exported header file is included by the C module realizing the inference engine for ADM-EDGE autoencoders prior to its integration into the firmware. The inference engine for ADM-FOG autoencoders is implemented in Python relying on the Tensorflow library. 

Decisions made by ADM-EDGE lightweight autoencoders are re-evaluated by ADM-FOG autoencoders in case of low confidence scores. The default value of the threshold is set to $C_{th}=0.75$, i.e., the decisions with $C<C_{th}$ are re-evaluated. We adopt here a standard, confidence-score based decision that is simple but effective; for more advanced mechanisms on how to offload decisions from the edge, see, e.g., \cite{Jaddoa_2020}. The threshold $C_{th}$ is a tunable parameter that allows to trade-off confidence in the decision about anomaly and response time. Lower threshold values correspond to the system designer's satisfaction with lower confidence scores, with the benefit that the average response time within a time interval for the same input data set is decreased.

\section{System Integration, Data Generation and Numerical Results}
\label{Sec4}

\subsection{System Integration}

To integrate the system, collect real-world data and perform testing and evaluation, CIoT UE is connected to the FGW via a mobile operator macro-cellular NB-IoT eNB. CIoT UE is running the ADM-EDGE software module and periodically sends data points to the FGW encapsulated into UDP packets. Within the mobile operator core network, a general purpose server is set and connected to the PGW gateway. The ADM-FOG software module within the server accepts UDP packets sent by CIoT UE. The server provides sufficient resources to run the ADM-FOG module, so in the sequel, we focus on the ADM-EDGE module deployment on the CIoT UE device.

To estimate the storage budget of an ADM-EDGE model with one hidden layer in terms of memory footprint the following results are given in Table~\ref{table_1}. One can note that ADM-EDGE consumes a small fraction of standard NB-IoT device firmware needed for basic device sensing, processing and communication functionality. Tensorflow and Tensorflow lite exported models sizes are also given for reference. Table~\ref{table_2} shows comparison of ADM-EDGE memory resource utilization for autoencoders with 1, 3 and 5 hidden layers (as in the previous table, the sizes of exported Tensorflow and Tensorflow light models are given for reference). It can be observed that additional hidden layers do not significantly increase the memory footprint of the ADM-EDGE module within the firmware of the edge node device.

Computational budget of ADM-EDGE devices is estimated by available number of operations per second. The ARM Cortex-M0+ ADM-EDGE CPU has a two-stage pipeline, and most instructions are executed within 1 clock cycle (some take 2 and a few take more than 2 clock cycles). The following holds for CPU: Peak throughput = Peak IPC * f = 1 * 16 MHz = 16 MOps/s, where peak IPC (instructions per cycle) = 1 for ARM Cortex M0+ architecture. Accordingly, peak computational budget is up to 16 MOps/s.

\begin{table}[thbp]
\caption{ADM-EDGE memory resource utilization for the ADM-EDGE autoencoder with one hidden layer.}
\begin{center}
\begin{tabular}{ll}
\textbf{MODEL} & \textbf{Size in bytes} \\
\noalign{\smallskip}\hline\noalign{\smallskip}
Firmware without ADM-EDGE & 55816 (21,3\%) out of 262144 \\
Firmware with ADM-EDGE &  61896 (23,6\%)  out of 262144 \\
ADM-EDGE only & 6080 (\url{~}2\%) \\
Tensorflow ADM & 21696 \\
Tensorflow lite ADM & 1452 \\
\noalign{\smallskip}\hline
\end{tabular}
\label{table_1}
\end{center}
\end{table}

\begin{table}[thbp]
\caption{ADM-EDGE memory resource utilization for ADM-EDGE autoencoders
with different number of hidden layers (HL). Sizes are given in bytes.}
\begin{center}
\begin{tabular}{llll}
\textbf{MODEL} & \textbf{1 HL} & \textbf{3 HL} & \textbf{5 HL} \\
\noalign{\smallskip}\hline\noalign{\smallskip}
ADM-EDGE only       & 6080  & 6524  & 8278  \\
Tensorflow ADM      & 21696 & 32240 & 44256 \\
Tensorflow lite ADM & 1452  & 2032  & 3056 \\
\noalign{\smallskip}\hline
\end{tabular}
\label{table_2}
\end{center}
\end{table}

\subsection{Data Generation}
\label{data_generation}
\begin{figure}[ht]
\centerline{\includegraphics[width=3.3in]{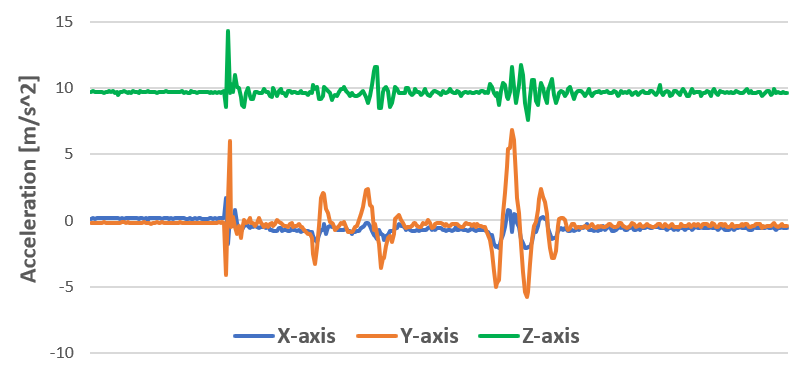}}
\caption{Example of acceleration data from IMU.}
\label{Fig_acceleration}
\end{figure}

To generate the dataset (elaborated in Section \ref{numerical_results}), we used NB-IoT edge nodes described in Section \ref{edge_device_design}. We created a setup where an edge node has been attached to a box-shaped container inside a transport vehicle moving through the city of Novi Sad. The device was initially connected to the NB-IoT network, and it had the uninterrupted connectivity throughout the path. We collected the positioning data from GNSS module (timestamp, latitude, longitude, altitude, speed and number of satellites in range), as well as the outputs of the IMU (acceleration and magnetic field along the 3 spatial axes). The time resolution (sampling period) of the GNSS samples was $\Delta_1=10$~s. The sampling period of the IMU is $\Delta_2=15$~ms (see Fig. \ref{Fig_acceleration} for an example of IMU signals), thus we calculated the RMS for the acceleration and magnetic field samples collected within a sampling interval $\Delta_1$ (as described in Sec. III.A). The collected data was stored at a database at the FGW, and it was used to train the AD model discussed in the following section.

\subsection{Numerical Results}
\label{numerical_results}

ADM-EDGE and ADM-FOG autoencoders are evaluated using two independent real-world datasets. The first dataset reflects the behaviour of the edge node device under normal driving conditions without large disturbances. This dataset contains 12678 data points and it is used to train ADM-EDGE and ADM-FOG autoencoders. The trained autoencoders are tested on the second dataset. The test dataset has 1571 data points with 42 intentionally caused anomalous events induced by shaking and overturning the container with the attached device. Since the edge node records both location-based features (GPS longitude and latitude) and IMU-based features, we can distinguish two types of anomalous events: location-based anomalies (large deviations from learned trajectories) and behaviour-based anomalies (large deviations from learned IMU signals). Our test dataset does not contain any location-based anomalies.

The accuracies of ADM-EDGE and ADM-FOG autoencoders are assessed by computing the following basic
measures:
\begin{itemize}
    \item $TP$ (true positives) -- the number of correctly identified anomalous events,
    \item $FP$ (false positives) -- the number of times an autoencoder indicated a non-existing anomalous event, and
    \item $FN$ (false negatives) -- the number of times an autoencoder missed to indicate an existing anomalous event.
\end{itemize}
We define the anomalous data points as those that correspond to the intentionally caused incident events; these data points are known to the experiment designer and system evaluator but are not known beforehand to the AD modules. The goal of AD is then to uncover the defined anomalies from the data in an unsupervised manner.

From $TP$, $FP$ and $FN$ we derive precision ($P$) and recall ($R$) scores of examined anomaly detection models:
\begin{equation}
\label{formula_precision}
P = \frac{TP}{TP + FP}
\end{equation}
\begin{equation}
\label{formula_recall}
R = \frac{TP}{TP + FN}
\end{equation}
Both precision and recall take values in the range [0, 1]. Precision indicates the degree of correctness of an anomaly detection model: small precision values imply that the model makes a lot of errors when stating anomalous events. Recall reflects the degree of model's ability to detect existing anomalous events. Small recall values indicate that the model often remains ``silent'' in cases when it should alarm anomalous events.

When comparing different anomaly detection models it is useful to have a single overall score reflecting their performances. For this purpose we use the $F_{1}$ measure which is the harmonic mean of precision and recall. The $F_{1}$ measure equally weights precision and recall favouring models that do not show extreme behaviour (high precision and low recall or vice versa):
\begin{equation}
\label{formula_F1}
F_{1} = \frac{2 \cdot P \cdot R}{P + R}.
\end{equation} 

In our experimental evaluation, we examine three ADM-EDGE anomaly detection
models (with 1, 3 and 5 hidden layers), 19 ADM-FOG models with three hidden layers (sequentially containing $n/2$, $n/4$ and $n/2$ nodes, where $n$ denotes the number of input features) accepting timeseries of lengths between $L=2$ to $L=20$, and 19 ADM-FOG models with five hidden layers (sequentially containing $3n/4$, $n/2$, $n/4$, $n/2$ and $3n/4$ nodes) also working with timeseries of lengths between $L=2$ and $L=20$. Due to the stochastic nature of the autoencoder learning algorithm, an ensemble of 20 autoencoders is trained for each examined model. All autoencoders are trained in maximally 200 epochs, with the batch size equal to 16 and the default learning rate of the Adam algorithm (0.001). Early stopping is activated after 10 epochs without a decrease in the value of the loss function.

The evaluation metrics for a particular model are estimated by averaging results individually obtained from all autoencoders in the corresponding ensemble. Additionally, two variants of each model are examined: a model trained without location-based features (NO-GPS case) and a model trained on all features (GPS case).

The performance of ADM-EDGE and ADM-FOG autoencoders is compared to five baseline anomaly detection methods that are not based on deep learning algorithms:
\begin{enumerate}
\item SVM -- anomaly detection using one-class support vector machines,
\item ABOD -- angle-based outlier detection,
\item KNN -- anomaly detection based on the $K$-nearest neighbours algorithm ($K = 10$),
\item PCA -- anomaly detection based on principal component analysis, and
\item HBOD -- histogram-based outlier detection.
\end{enumerate}
The implementation of baseline methods from the PyOD anomaly detection library~\cite{PYOD2019} is used in our comparative analysis with default values for hyper-parameters (as specified in PyOD). Regarding the 3GPP CIoT architecture augmented with anomaly detection enhancements, baseline anomaly detection methods trained and evaluated on individual data points map to the edge layer, while the same baselines trained and tested on timeseries formed from consecutive data points correspond to the fog layer.

The results of the evaluation of ADM-EDGE autoencoders in both variants (GPS and NO-GPS case) are summarized in Table~\ref{table_edge}. The table also shows precision, recall and $F_{1}$ scores for baseline methods when trained and tested on individual data points (i.e., in the same setting as ADM-EDGE autoencoders). It can be observed that ADM-EDGE autoencoders have considerably balanced precision and recall scores: the largest difference between precision and recall is equal to 0.096 (ADM-EDGE-3 in the NO-GPS case). Both scores are slightly varying around 0.7 implying that ADM-EDGE autoencoders exhibit a quite good performance. There are no large differences in obtained precision and recall scores in GPS and NO-GPS cases, which is also evident by similar values of $F_{1}$ scores. This result is expected since the test dataset does not contain location-based anomalies. Therefore, small differences in obtained results can be explained by the stochastic nature of the autoencoder learning algorithm. The largest $F_{1}$ score in both cases is achieved by the ADM-EDGE autoencoder with 5 hidden layers. However, ADM-EDGE autoencoders with 1 and 3 hidden layers have slightly lower $F_{1}$ scores (especially in the NO-GPS case) implying that they are equally effective anomaly detection models. 

\begin{table}[htb!]
\caption{Evaluation of ADM-EDGE autoencoders and comparison to baseline methods (GPS -- location-based features included, NO-GPS -- location-based features excluded). ADM-EDGE-$k$ denotes the ADM-EDGE autoencoder with $k$ hidden layers.}
\begin{center}
\begin{tabular}{lllllll}
\noalign{\smallskip}\hline \noalign{\smallskip}
 & \multicolumn{3}{l}{\textbf{GPS}} & \multicolumn{3}{l}{\textbf{NO-GPS}} \\
\noalign{\smallskip}\hline \noalign{\smallskip}
 & $P$ & $R$ & $F_{1}$ & $P$ & $R$ & $F_{1}$ \\
 \noalign{\smallskip}\hline \noalign{\smallskip}
ADM-EDGE-1 &	0.705	& 0.690	& 0.697	& 0.710	& 0.676	& 0.693\\
ADM-EDGE-3 &	0.706	& 0.697	& 0.702	& 0.739	& 0.644	& 0.688\\
ADM-EDGE-5 &	0.681	& 0.764	& 0.720	& 0.701	& 0.697	& 0.699\\
\noalign{\smallskip}\hline \noalign{\smallskip}
SVM	 & 0.467	& 0.179	& 0.259	& 0.467	& 0.179	& 0.259 \\
ABOD & 	0.488	& 1.000	& 0.655	& 0.488	& 1.000	& 0.655 \\
KNN	 & 0.500	& 0.692	& 0.581	& 0.500	& 0.692	& 0.581 \\
PCA	 & 0.800	& 0.410	& 0.542	& 0.714	& 0.513	& 0.597 \\
HBOD & 1.000	& 0.077	& 0.143	& 0.654	& 0.436	& 0.523\\
\noalign{\smallskip}\hline \noalign{\smallskip}
\end{tabular}
\label{table_edge}
\end{center}
\end{table}

The results presented in Table~\ref{table_edge} also show that ADM-EDGE autoencoders perform better than the baseline anomaly detection methods:
all ADM-EDGE autoencoders have higher $F_{1}$ scores compared to baselines. The most successful baseline method considering $F_{1}$ is ABOD. ABOD indicates the same set of anomalous events in both cases with the highest possible recall (no false negatives), but its precision is less than 0.5, implying that ABOD makes a lot of false positive anomaly decisions. KNN and PCA have a significantly better balance between precision and recall compared to ABOD, but at a lower level of the $F_{1}$ score. Two worst performing baselines are SVM and HBOD. SVM has a very low recall in both cases with precision less than 0.5 resulting with a low $F_{1}$ score around 0.25. In the GPS case, HBOD exhibits the highest possible precision (no false positives) but with an extremely low recall (less than 0.1). Much more balanced HBOD scores are present in the NO-GPS case indicating that HBOD is highly sensitive to the addition of GPS features.

\begin{figure}[htb!]
\centerline{\includegraphics[scale=0.4]{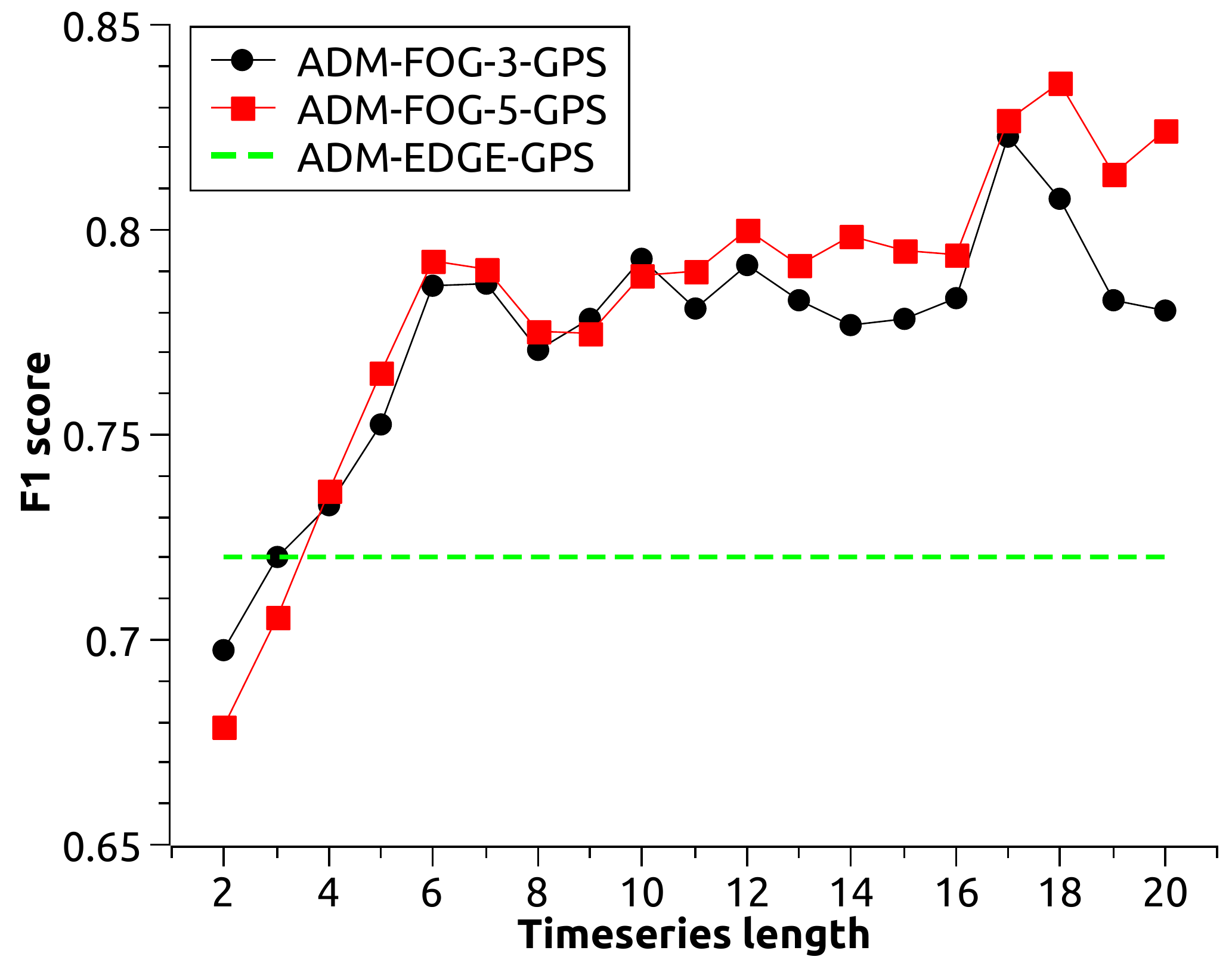}}
\caption{$F_{1}$ scores of ADM-FOG autoencoders (GPS features included) with 3 and 5 hidden layers for different timeseries lengths. The dashed line indicates the F1 score of the best ADM-EDGE autoencoder with GPS features included.}
\label{fig_1}
\end{figure}

\begin{figure}[htb!]
\centerline{\includegraphics[scale=0.4]{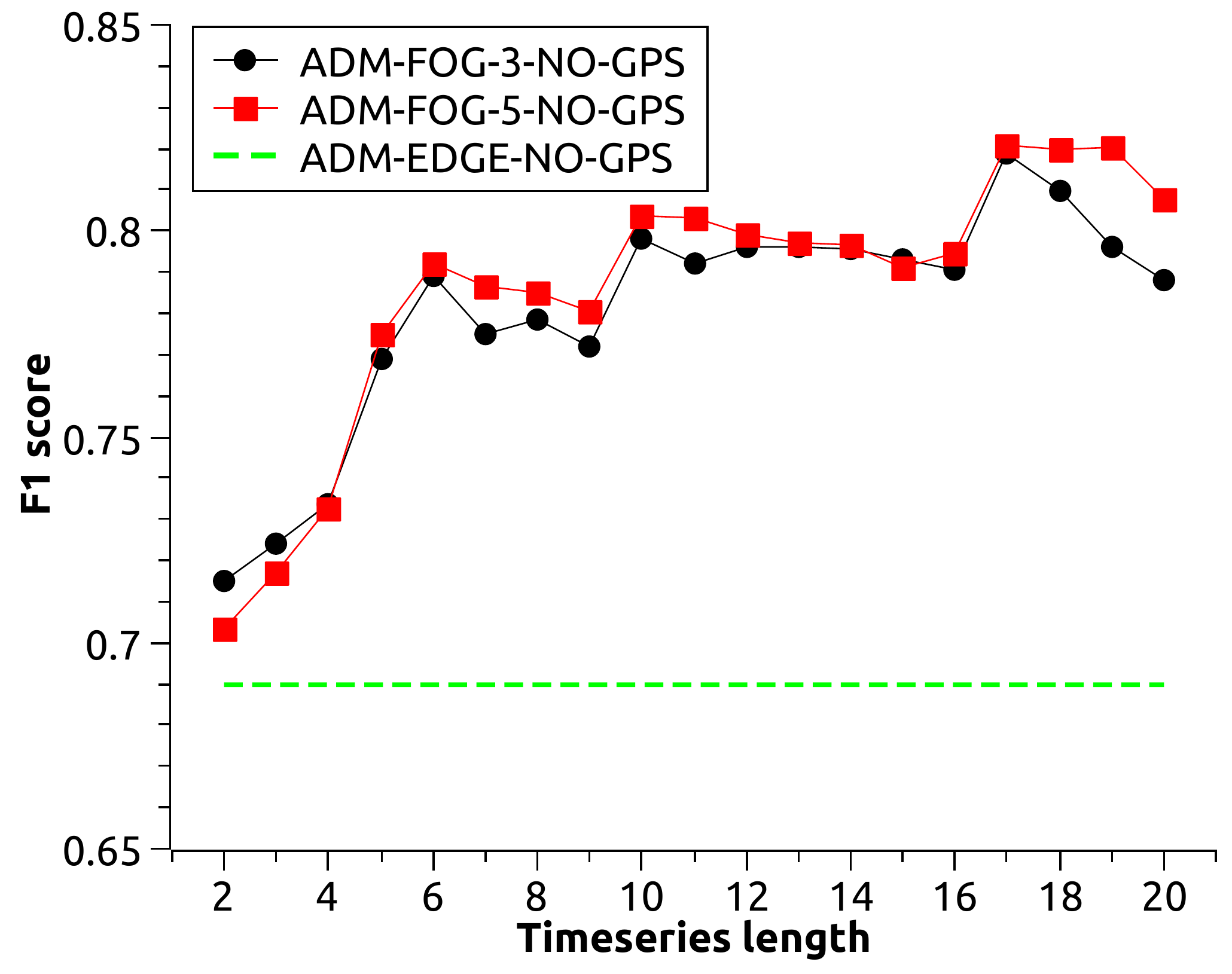}}
\caption{$F_{1}$ scores of ADM-FOG autoencoders (GPS features excluded) with 3 and 5 hidden layers for different timeseries lengths. The dashed line indicates the F1 score of the best ADM-EDGE autoencoder with GPS features excluded.}
\label{fig_2}
\end{figure}

In the second experiment we examine the performance of ADM-FOG autoencoders with 3 and 5 hidden layers. The obtained F1 scores are presented in Figure~\ref{fig_1} for the GPS case and in Figure~\ref{fig_2} for the NO-GPS case. It can be seen that in both cases ADM-FOG autoencoders with 3 hidden layers achieve $F_{1}$ scores close to those of ADM-FOG autoencoders with 5 hidden layers. Similarly as for ADM-EDGE autoencoders, the location-based features do not have a significant impact to the performance of ADM-FOG autoencoders: the largest difference in $F_{1}$ scores across cases is equal to 0.044. ADM-FOG autoencoders trained on all features (GPS case) exhibit higher $F_{1}$ scores than the best ADM-EDGE autoencoder when timeseries are longer than 3 data points. The average increase in the F1 score when offloading anomaly detection decisions to ADM-FOG is 7.36\%. In the NO-GPS case, ADM-FOG autoencoders achieve higher $F_{1}$ scores than the best ADM-EDGE autoencoder through the whole range of timeseries lengths. The average improvement in $F_{1}$ in this case is equal to 9.56\%.

The results shown in Figures~\ref{fig_1} and~\ref{fig_2} allow us to explicitly quantify trade-offs between performance of anomaly detection and response time, with respect to whether the decision on the presence of anomalies is carried out at the edge or at the fog. For this, note that the response time of ADM-EDGE corresponds approximately to one sampling period $\Delta_1$. On the other hand, the response time of ADM-FOG depends on the length $L$ of the timeseries processed. In the case of ADM-FOG autoencoders trained without location-based features (NO-GPS case), the largest $F_{1}$ score is achieved by the autoencoder with 5 hidden layers working on timeseries of length $L=17$ ($P = 0.821$, $R = 0.8205$, $F_{1} = 0.8208$). The increase in precision and recall compared to the best ADM-EDGE autoencoder is equal to 0.12 and 0.123, respectively. This means that by increasing the confidence threshold for offloading anomaly detection decisions to the ADM-FOG autoencoder the whole system has significantly less both false negative and false positive decisions at the cost of decision delays by $L=17$ time slots. The ADM-FOG autoencoder with 5 hidden layers working on timeseries of length $L=18$ has the highest $F_{1}$ scores among FOG models trained on all features ($P = 0.8255$, $R = 0.8462$, $F_{1} = 0.8357$). The increase in precision and recall in this case is 0.14 and 0.08, respectively. Therefore, by increasing the offloading threshold the performance of the whole system improves by having less false negative decisions and significantly less false positive decisions at the cost of decision delays by $L=18$ time slots.

\begin{figure}[htb!]
\centerline{\includegraphics[scale=0.4]{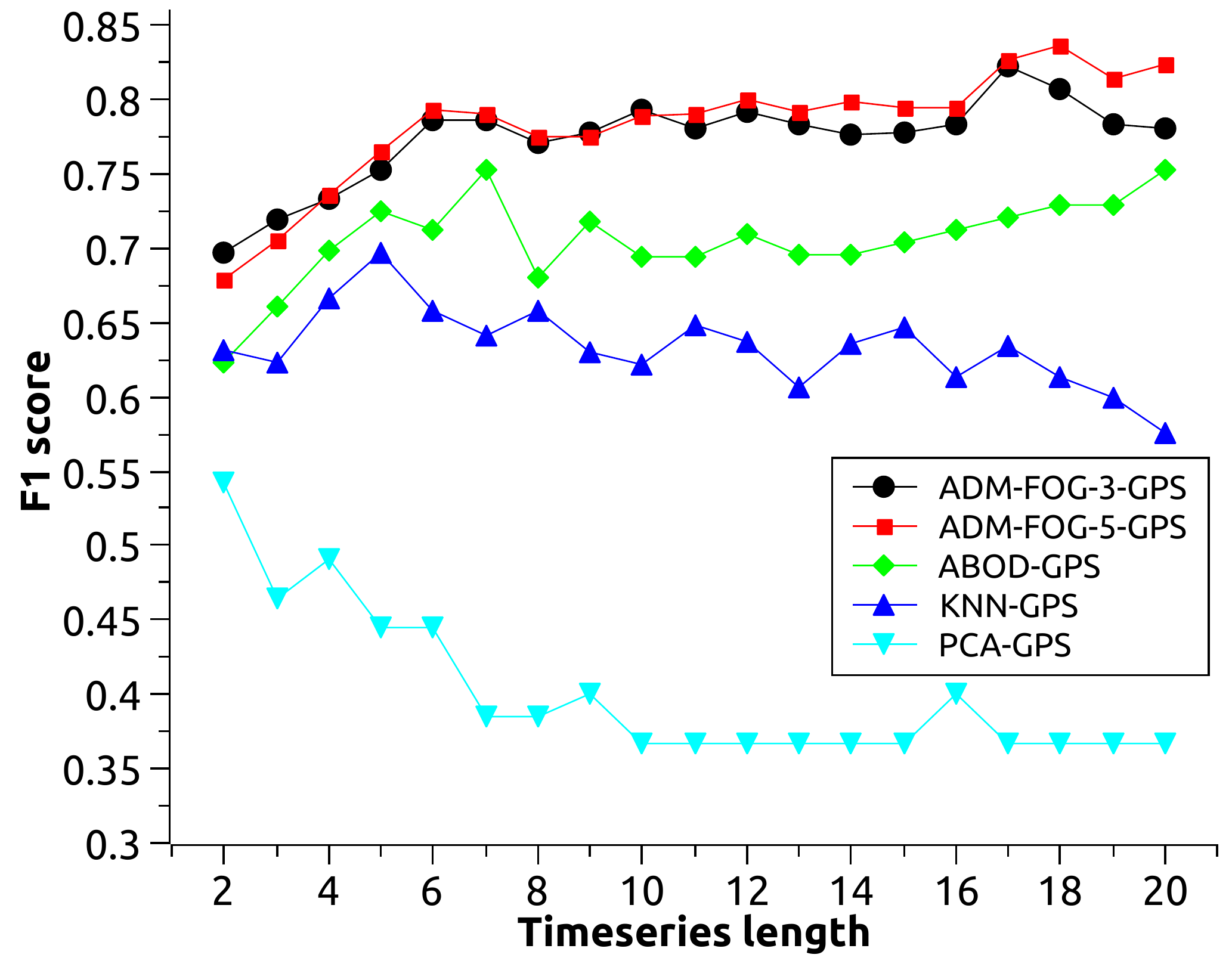}}
\caption{Comparison of $F_{1}$ scores of ADM-FOG autoencoders with GPS features included to $F_{1}$ scores of baseline methods for different timeseries lengths. SVM and HBOD have $F_{1}$ scores less than 0.3 through the whole range of timeseries lengths.}
\label{fig_3}
\end{figure}

\begin{figure}[htb!]
\centerline{\includegraphics[scale=0.4]{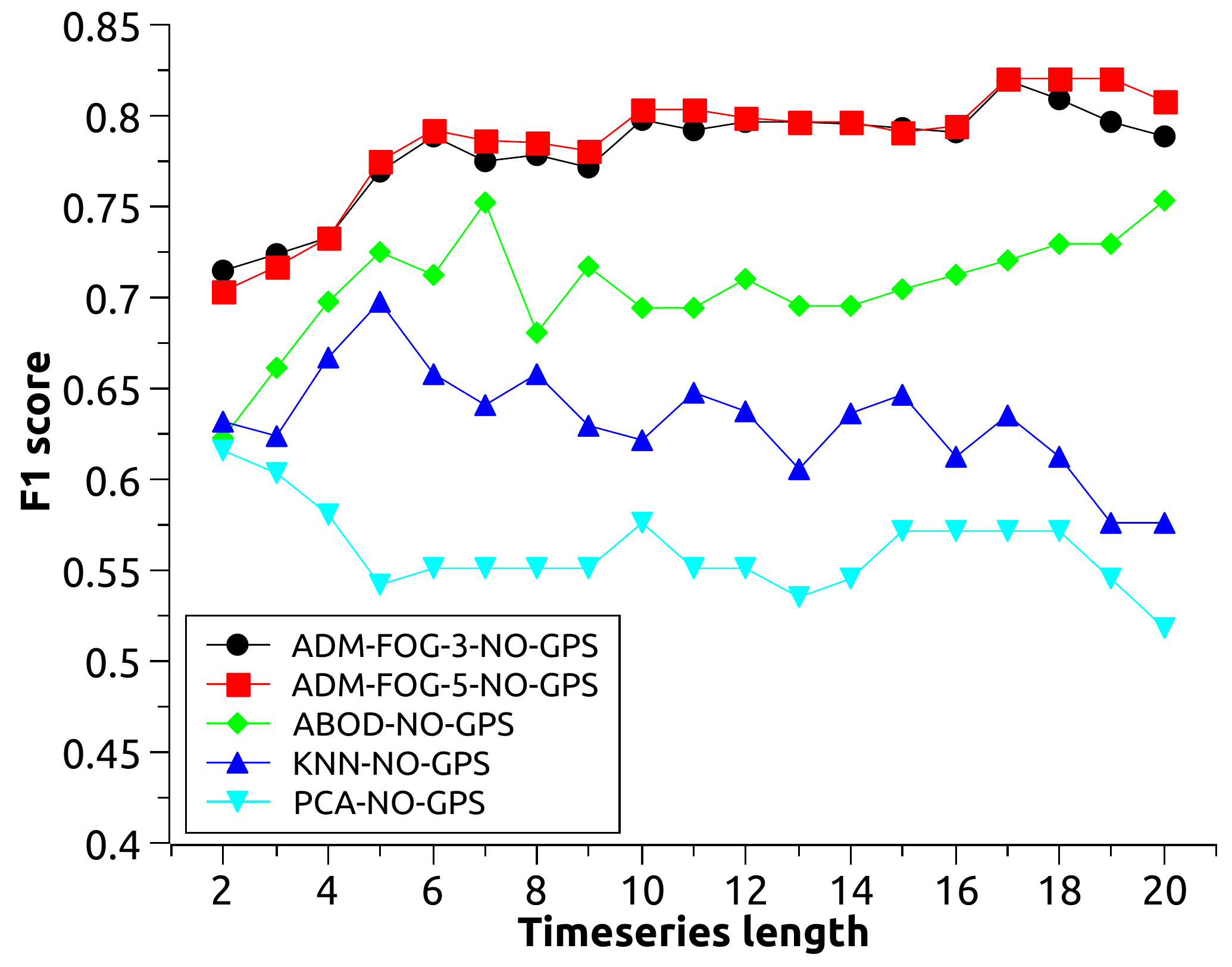}}
\caption{Comparison of $F_{1}$ scores of ADM-FOG autoencoders with GSP features excluded to $F_{1}$ scores of baseline methods for different timeseries lengths. SVM and HBOD have $F_{1}$ scores less than 0.45 through the whole range of timeseries lengths.}
\label{fig_4}
\end{figure}

In the last experiment, we compare ADM-FOG autoencoders to baseline anomaly detection methods trained and tested on timeseries. The obtained $F_{1}$ scores for baselines are shown in Figure~\ref{fig_3} (GPS case) and Figure~\ref{fig_4} (NO-GPS case), together with $F_{1}$ scores of ADM-FOG autoencoders for comparison. SVM and HBOD that are two worst performing baseline methods on individual data points are also the worst performing baselines on timeseries and they are not shown in Figures~\ref{fig_3} and~\ref{fig_4} ($F_{1}$ scores less than 0.3 and 0.45 for GPS and NO-GPS case, respectively, through the whole range of timseries lengths). It can be seen that for an arbitrary timseries length ADM-FOG autoencoders achieve higher $F_{1}$ scores compared to baselines in both cases. The ranking of methods according to their $F_{1}$ scores in
the GPS case is the following:
\begin{equation*}
\label{method_ranking_1}
\text{ADM-FOG} \succ \text{ABOD} \succ \text{KNN} \succ \text{PCA} \succ \text{SVM}
\succ \text{HBO},
\end{equation*}
where $\succ$ means "performs better than". For the NO-GPS case we have:
\begin{equation*}
\label{method_ranking_2}
\text{ADM-FOG} \succ \text{ABOD} \succ \text{KNN} \succ \text{PCA} \succ \text{HBO}
\succ \text{SVM}.
\end{equation*}

Our experimental evaluation shows that both ADM-EDGE and ADM-FOG autoencoders perform better than all examined baseline methods. Therefore, it can be concluded that autoencoders are an adequate choice to enhance the proposed CIoT architecture with unsupervised anomaly detection capabilities at both edge and fog layer.

\section{Conclusion}
\label{Sec5}

In this paper, we presented design, implementation, real-world deployment and evaluation of a novel anomaly detection architecture for Cellular IoT networks tailored for the Smart Logistics use case. We demonstrate and quantify major system-design trade-offs between responsiveness and accuracy with respect to the position (i.e., edge or fog) within the Cellular IoT network where anomaly detection is performed. Through real-world deployment study, we emphasize that autoencoders represent a suitable choice for ML anomaly detection at the edge. 

The results reported in this paper are based on a small-scale real-world trial. For our future work, the trial will be extended to a large number (approximately 50) container-carrying vehicles in a realistic nation-wide Smart Logistics use case. Possibilities to perform training process at the edge devices will be explored, as well as opportunities to integrate advanced distributed learning concepts such as federated learning as part of the proposed Cellular IoT architecture will be investigated.



\ifCLASSOPTIONcaptionsoff
  \newpage
\fi

\end{document}